# Structural, optical, and electrical properties of Cu-doped NiO films synthesized by spray pyrolysis for potential gas sensing applications

**Eka Nurfani[1,2,*], Grace Grace[3], Mahardika Yoga Darmawan[3], Resti Marlina[4], Jumaeda Jatmika[5], Asnan Rinovian[6], Aditya Rianjanu[1,2]**

[1]Department of Materials Engineering, Faculty of Industrial Technology, Institut Teknologi Sumatera (ITERA), Lampung 35365, Indonesia.

[2]Center for Green and Sustainable Materials, Institut Teknologi Sumatera (ITERA), South Lampung 35365, Indonesia

[3]Department of Physics, Faculty of Science, Institut Teknologi Sumatera (ITERA), Lampung 35365, Indonesia.

[4]Research Center for Biomass and Bioproducts, National Research and Innovation Agency (BRIN), Cibinong 16911, Indonesia

[5]Research Center for Quantum Physics, National Research and Innovation Agency (BRIN), South Tangerang 15314, Indonesia

[6]Research Center for Mining Technology, National Research and Innovation Agency (BRIN), Lampung 35361, Indonesia

*Corresponding author: eka.nurfani@mt.itera.ac.id

## Abstract

Cu-doped NiO thin films were deposited on ITO substrates via spray pyrolysis at 450 °C for 1.5 minutes. XRD confirmed a cubic NiO structure, with Cu incorporation reducing crystallite size, increasing interplanar spacing, and expanding the lattice parameter, indicating successful substitution of Cu ions. Optical analysis showed a slight bandgap reduction from 3.70 to 3.65 eV, while PL revealed lower emission intensity, suggesting enhanced defect states and suppressed carrier recombination. Raman spectroscopy exhibited redshifts in LO, 2TO, and 2LO modes and the disappearance of the TO mode, confirming lattice distortion from Cu doping. Gas sensing tests under ambient and LPG conditions demonstrated significantly improved sensitivity and voltage-dependent response for doped films. These results establish that Cu incorporation enhances charge transport and gas interaction mechanisms, making Cu-doped NiO films highly promising for efficient and reliable gas sensors.

## 1 Introduction

The rapid advancement of technology has spurred the development of innovative gas sensor systems, particularly those based on semiconductor materials. Semiconductor-based gas sensors offer high sensitivity, fast response times, and simple integration into electronic devices, making them attractive for detecting hazardous gases such as LPG, CO, and $NO_2$ [1]. Metal oxide semiconductors, such as ZnO [2], $SnO_2$ [3], $TiO_2$ [4], and NiO [5], are commonly used as sensing layers due to their ability to interact with gas molecules on their surfaces, resulting in measurable changes in electrical properties. These interactions are highly dependent on surface activity, making material composition and structure critical to performance.

Among these oxides, nickel oxide (NiO) has gained significant attention as a gas-sensing material due to its chemical stability, low cost, and favorable electrical and optical properties [5,6]. As a p-type semiconductor with a cubic crystal structure, NiO is especially promising for detecting reducing gases. However, its relatively wide bandgap (~3.7 eV) and limited carrier mobility hinder its sensitivity and response at room temperature [7]. To address these limitations, researchers have explored doping NiO with various transition and post-transition metals, such as Cu [8], Co [9], Al [10], and Fe [11], to introduce defect states, modify the band structure, and enhance surface reactivity. Among them, copper (Cu) stands out due to its high electrical conductivity and compatibility with the crystal lattice of NiO, offering potential improvements in both charge transport and gas sensitivity.

Furthermore, spray pyrolysis has emerged as a simple, cost-effective, and scalable technique for fabricating metal oxide thin films [12]. Unlike other deposition methods, such as physical vapor deposition [13], spray pyrolysis does not require expensive precursors, vacuum systems, or post-deposition annealing. Its ultra-fast synthesis time—allowing film growth within minutes—makes it particularly attractive for rapid prototyping and large-scale production [14,15]. Moreover, the technique is compatible with a variety of substrates and is well-suited for doping processes, making it ideal for developing advanced gas sensor materials [16]. Despite its advantages, the application of ultra-short deposition spray pyrolysis (≤2 minutes) for Cu-doped NiO gas sensors remains underexplored, highlighting a gap in current research.

In this study, Cu-doped NiO thin films were successfully synthesized using an ultra-fast spray pyrolysis technique with a deposition time of only 1.5 minutes. Structural, optical, and electrical characterizations were conducted to investigate the effect of Cu doping on the properties of NiO films. Furthermore, gas sensing performance was evaluated against LPG gas to determine sensitivity enhancement due to doping. The novelty of this research lies in the combination of ultra-fast deposition and Cu doping for optimizing NiO-based gas sensors, providing a promising strategy for low-cost and efficient gas detection technologies.

## 2 Methods

NiO and NiO:Cu thin films were grown on an ITO-coated glass substrate using spray pyrolysis. For the pure NiO, a total of 0,1 M nickel chloride hexahydrate (Merck, for analysis) was dissolved into 40 mL of distilled water and then stirred using a magnetic stirrer for 1 hour at room temperature. For the NiO:Cu sample, 7 wt% of copper chloride dihydrate (Merck, for analysis) was added relative to the nickel chloride hexahydrate, with similar molarity and volume. The reaction for the formation of the primary precursor is described in **Equation (1)**:

$$\begin{gathered} NiCl_2{\cdot}6H_2O{+}H_2O \rightarrow NiO{+}2HCl{+}6H_2O \\ CuCl_2{\cdot}2H_2O \rightarrow Cu(H_2O)_2{+}Cl_2 \\ NiO{+}Cu(H_2O)_2 \rightarrow NiOCu{+}2H_2O \end{gathered} \tag{1}$$

Next, the ITO substrate was washed using an ultrasonic cleaner with an ethanol solution and distilled water for 10 minutes each. After cleaning the substrate, the hotplate temperature was raised to 450 ℃ for approximately 15 minutes to achieve a homogeneous temperature on the substrate. The precursor poured into the spray container was sprayed onto the surface of the ITO substrate [17]. The deposition was conducted for 1.5 minutes.

Furthermore, the structural properties of the thin films were characterized using X-ray diffraction (XRD, Bruker D8 Advance) and field-emission scanning electron microscopy (FESEM, Quattro S) to determine their crystal structures and surface morphology. UV-Vis (Bruker D8 Advance) spectroscopy was employed to measure the optical properties. Raman and photoluminescence (PL) spectroscopy (excitation wavelength of 532 nm, 600 gratings) were used to characterize the vibrational modes after introducing the dopant and determine the optical properties and defects of the material. I-V characterization with planar configuration was used to assess the electrical properties and their sensitivity to gases. The concentration of liquefied petroleum gas (LPG) was around 1 vol%, composed mainly of propane and butane. Details of the schematic experiment were illustrated in **Figure 1**.

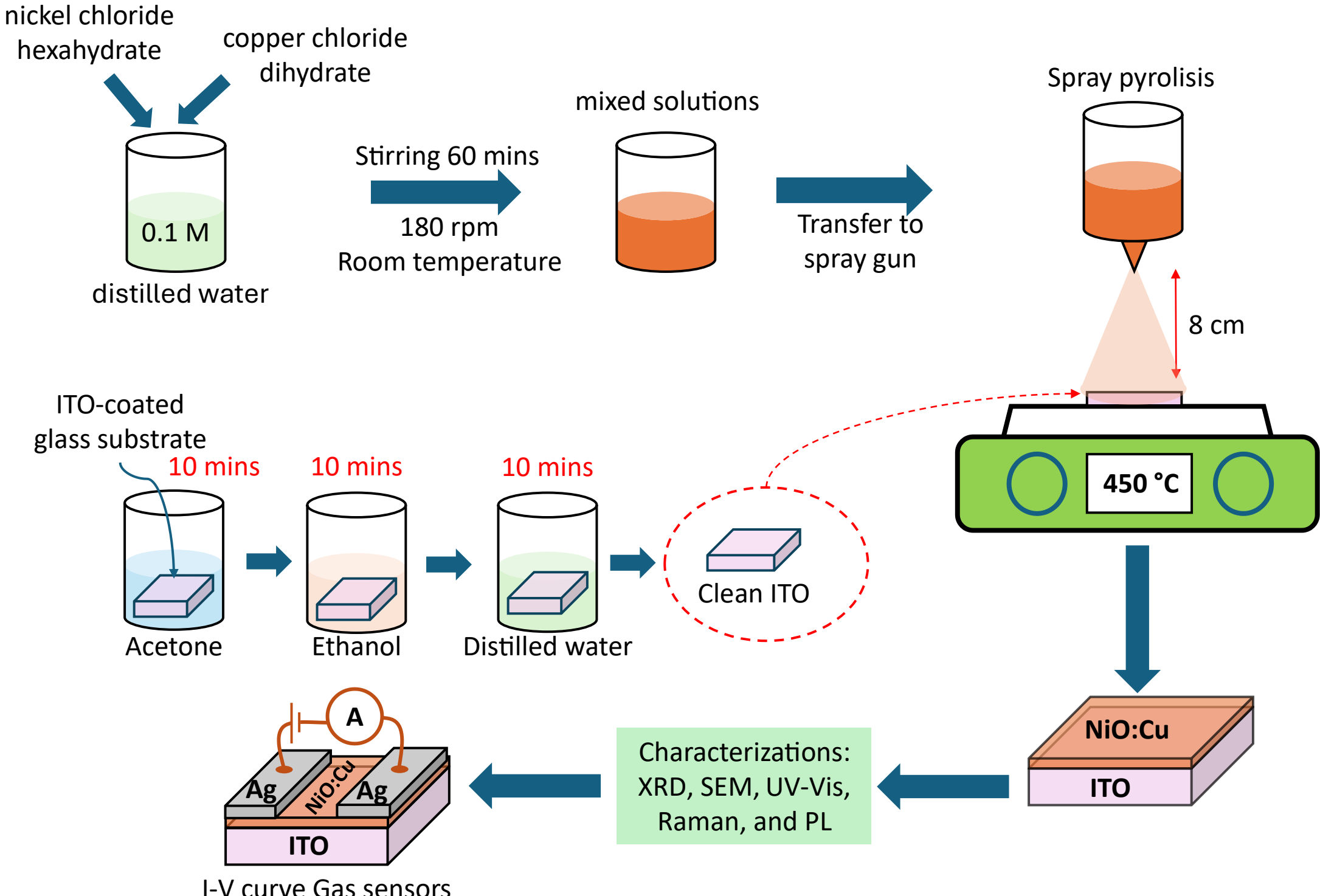

**Figure 1. Schematic experiment of NiO:Cu film deposited on the ITO-coated glass substrate and its characterizations.**

## 3 Results and Discussion

### 3.1 Structural Properties

**Figure 2(a)** compares the diffraction pattern of NiO and NiO:Cu thin films. An identified peak at 37° is identical to the NiO peak, as confirmed by the Crystallography Open Database (COD) reference number 96-432-0500 [18]. Based on the database, the identified NiO phase is cubic, with the measured Miller index (hkl) value of (111). The identified NiO peak is at 37.2°, while the identified NiO:Cu peak is at 37.0°. The peaks identified on the graph of the NiO and NiO:Cu thin films are dominated by peaks originating from the ITO-coated substrate. The X-ray waves can penetrate the thin film to the ITO substrate, as previously reported [19]. Based on this spectrum, the Cu diffraction peak is not identified because the $Cu^{2+}$ atom occupies the position of the $Ni^{2+}$ atom without changing the NiO crystal structure. So, it can be interpreted that the doping was successful.

From the XRD pattern, quantitative analysis can also be carried out in terms of interplanar spacing, crystallite size, and lattice parameters. The distance between crystals is based on Bragg's law, which is explained by **Equation (2)**,

$$2d \sin\theta = n\lambda \tag{2}$$

where $n$ is an integer as the order of refraction, $\lambda$ is the X-ray wavelength (1.54056 Å), $d$ is the distance between the two crystal planes, and $\theta$ is the angle between the incident ray and the normal plane. The obtained $d$ values are 2.41 and 2.42 Å for NiO and NiO:Cu films. The lattice parameters can be calculated according to **Equation (3)**:

$$\frac{1}{d^2} = (h^2 + k^2 + l^2)\frac{1}{a^2} \tag{3}$$

The obtained lattice parameters $(a = b = c)$ are 4.18 and 4.20 Å for NiO and NiO:Cu films. The increase of lattice parameter is due to a different ionic radius. $Ni^{2+}$ (0.78 Å) as the host has smaller ionic radii as compared to $Cu^{2+}$ (0.82 Å), making the crystal lattice expansion [20]. The crystallite size (*D*) value can be calculated using the Debye-Scherrer formula, as expressed by **Equation (4)**.

$$\boldsymbol{D} = \frac{\boldsymbol{K\,\lambda}}{\boldsymbol{\beta\ \cos\theta}} \tag{4}$$

where *D* is the crystallite size, *K* is the form factor of the crystal, and $\lambda$ is the X-ray wavelength, and β is the Full Width at Half Maximum (FWHM). The calculated crystallite size is decrease from 21 to 20 nm after introducing Cu dopant. The data from the quantitative analysis of the XRD characterization are presented in **Table 1.**

**Figure 2(b)** presents the scanning electron microscopy (SEM) image of the NiO thin film deposited on the ITO-coated glass substrate. The surface morphology reveals well-dispersed crystalline grains across the film surface. The particles exhibit a variety of sizes, ranging from submicron to a few microns, with mostly cuboidal or polyhedral shapes, indicating a well-developed crystalline growth. The uniform distribution of particles suggests that the deposition process resulted in a homogeneous coverage of the substrate. The presence of relatively large and distinct grains may indicate a preferred crystal growth orientation, consistent with the NiO (111) diffraction peak observed in the X-ray diffraction (XRD) pattern. Additionally, some smoother regions are visible between the crystalline grains, which may correspond to amorphous areas or thin film layers that are not fully crystallized. The scale bar of 5 μm shows that the NiO grains have a high surface density with small inter-grain distances, potentially facilitating efficient charge transport paths. This morphology is favorable for applications such as electronic devices, photocatalysis, or energy storage, where good crystallinity and uniform surface coverage are crucial.

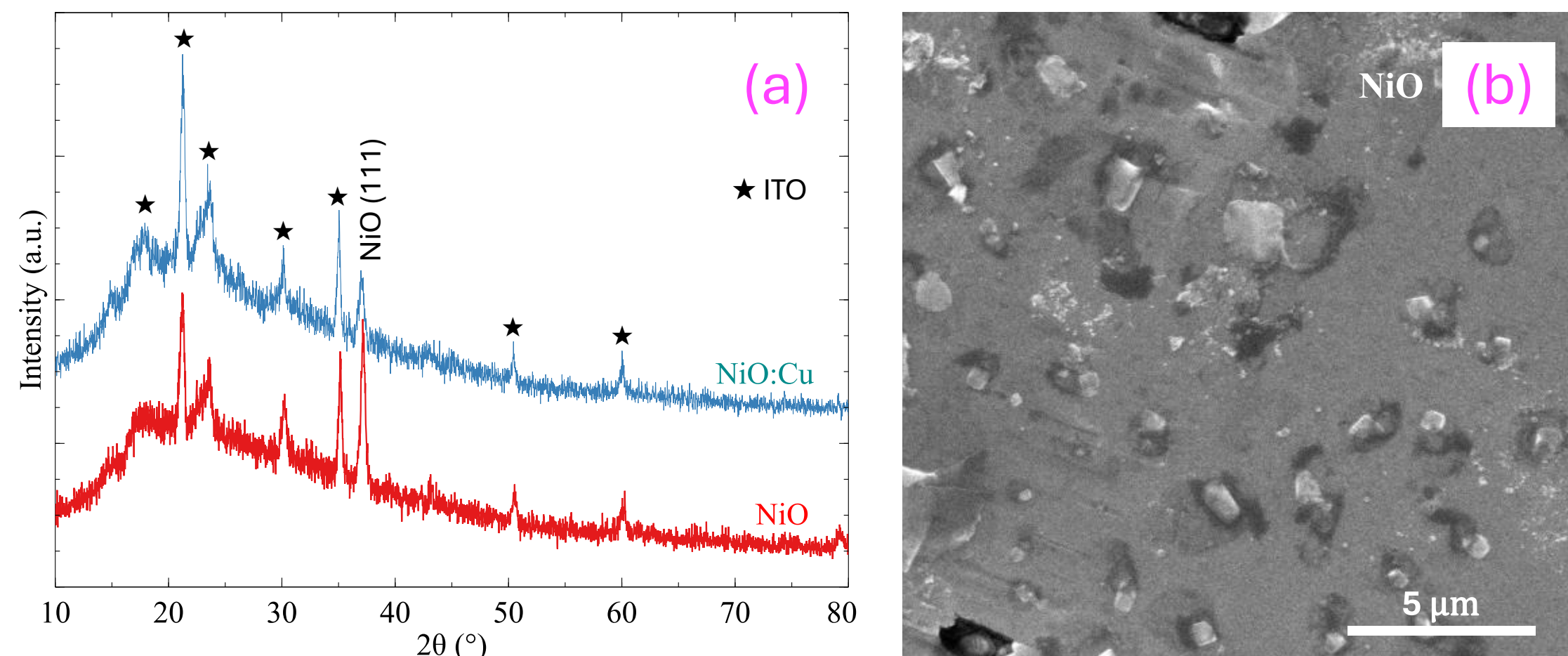

**Figure 2. (a) XRD pattern of pure NiO and NiO:Cu films on ITO-coated glass substrate. (b) Morphological images of NiO films captured by FESEM apparatus.**

**Table 1. Summary of XRD analysis including crystallite size, interplanar spacing, and lattice parameters.**

| Samples | $2\theta$ (°) | $d$ (Å) | $a=b=c$ (Å) | $D$ (nm) |
|---|---|---|---|---|
| NiO | 37.2 | 2.41 | 4.18 | 21 |
| NiO:Cu | 37.0 | 2.42 | 4.20 | 20 |

### 3.2 Optical Properties

The optical absorption and transmission characteristics of undoped and Cu-doped NiO films deposited via spray pyrolysis were systematically investigated through UV-Vis spectroscopy (**Figure 3**). Both films exhibit strong absorption in the ultraviolet region (<400 nm), consistent with the fundamental band-edge absorption of NiO. The NiO sample exhibits a higher absorbance value at wavelengths of 300-350 nm compared to NiO:Cu. Crucially, the Cu-doped film (NiO:Cu) demonstrates significantly enhanced absorbance across the visible spectrum (400–800 nm) compared to undoped NiO. This broadband absorption enhancement indicates the introduction of sub-bandgap states through Cu incorporation, likely arising from defect-mediated transitions (e.g., oxygen vacancies or Cu-related centers) and free-carrier absorption due to increased hole concentration from acceptor doping [21,22]. Such behavior is also found in other metal oxide semiconductors such as ZnO, $SnO_2$, and $TiO_2$.

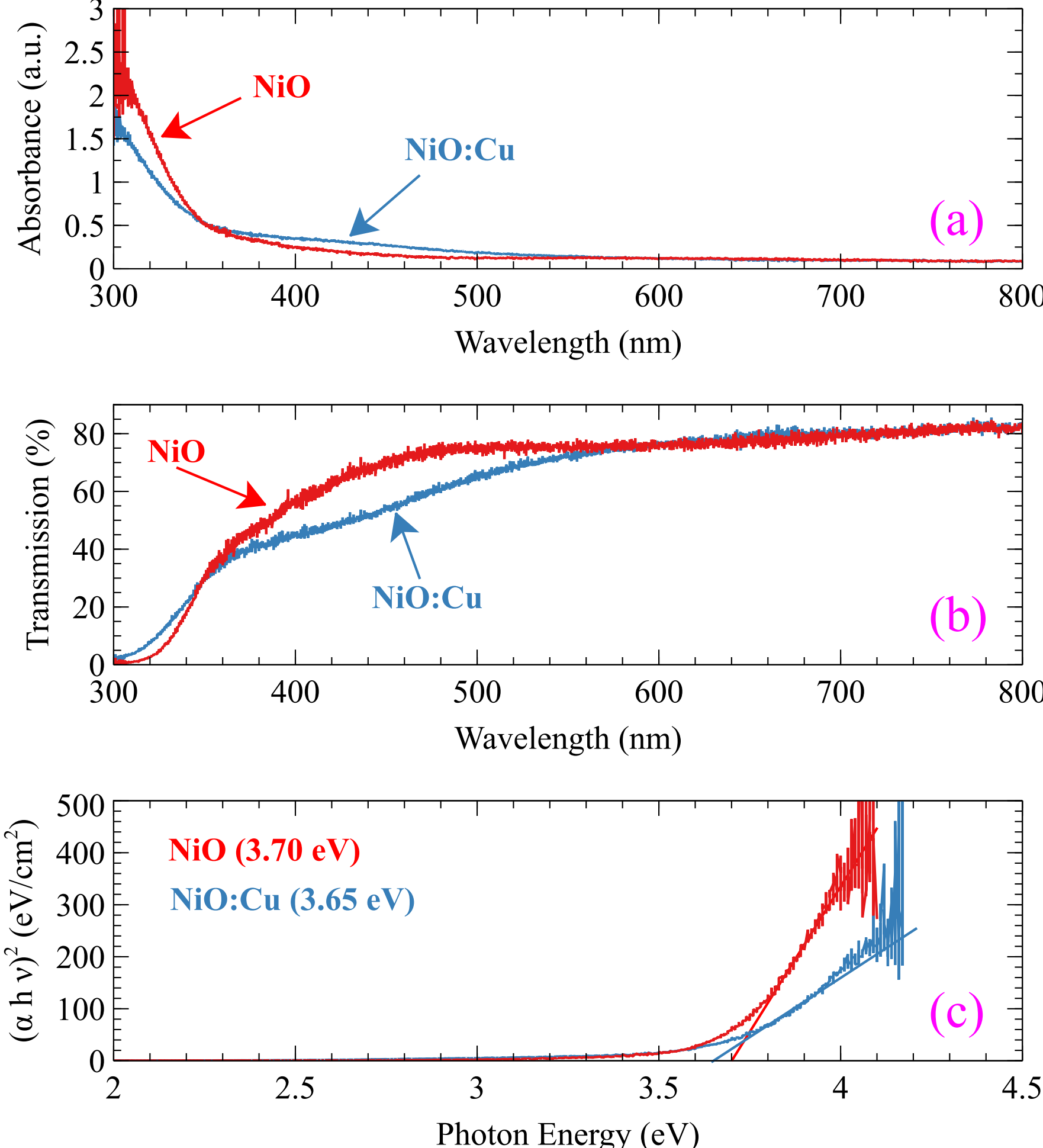

**Figure 3. (a) Absorbance and (b) transmittance spectra of NiO and NiO:Cu films grown on ITO-coated glass substrates. (c) Tauc-plot method to calculate the optical bandgap of NiO and NiO:Cu films.**

The bandgap is determined using the Tauc-plot method, which involves linearly extrapolating the curve with the longest gradient (slope). This extrapolation is carried out on the photon energy value on the X-axis. The relationship between the coefficient of absorption ($\alpha$) and photon energy can be seen according to **Equation (5)**:

$$(\alpha h\nu)^{\frac{1}{n}} = b\,(h\nu - E_g) \tag{5}$$

where α is the absorption coefficient, $h$ is Planck's constant (J.s), $\nu$ is the frequency (Hz), $b$ is a constant, $E_g$ is the bandgap energy (eV), and the component $n$ depends on the type of transition. It could be $n = 0.5$ for direct-allowed transition and $n = 2$ for indirect-allowed transition. In **Figure 3c,** the Tauc plot analysis reveals a measurable reduction in the direct band gap ($E_g$) from 3.70 eV (NiO) to 3.65 eV (NiO:Cu), confirming successful electronic structure modification. This 0.05 eV narrowing ($\Delta E_g$) is attributed to multiple synergistic mechanisms: (i) lattice strain induced by ionic radius mismatch between $Cu^{2+}$ (0.82 Å) and host $Ni^{2+}$ (0.78 Å) as evidenced by XRD peak shifts; (ii) formation of impurity bands near the valence band edge from Cu-3d orbitals; and (iii) band tailing associated with increased oxygen vacancy concentration. The reduced band gap aligns with the observed elevation in visible-light absorption, as sub-gap states enable lower-energy photon excitation. The red shift of the optical bandgap after Cu doping agrees with with previous results [23–25].

This band gap engineering has profound implications for gas sensing functionality. The narrowed $E_g$ facilitates thermal generation of holes, which are the primary charge carriers in p-type NiO, thereby enhancing baseline conductivity, as corroborated by I-V measurements. Concurrently, defect-mediated absorption indicates a high density of surface-active sites, particularly oxygen vacancies, which serve as adsorption centers for atmospheric oxygen species ($O_2^-$, $O^-$). During gas sensing events, these sites catalyze charge transfer with target analytes, thereby modulating the hole concentration in the valence band. The synergy of improved charge transport (from reduced $E_g$) and abundant adsorption sites, Cu-doped NiO is a promising candidate for high-sensitivity chemiresistive sensors, where surface redox reactions govern conductivity switching kinetics.

Moreover, Raman spectroscopy was performed to evaluate the vibrational modes and structural modifications induced by Cu doping in NiO thin films, as shown in **Figure 4a**. For the undoped NiO, characteristic peaks were observed at 493 $cm^{-1}$ (TO), 577 $cm^{-1}$ (LO), 776 $cm^{-1}$ (2TO), 988 $cm^{-1}$ (TO+LO), and 1089 $cm^{-1}$ (2LO), consistent with the phonon modes of cubic NiO [8,26–28]. Upon Cu doping, the TO mode disappeared, indicating possible local structural disorder or suppression due to symmetry breaking from $Cu^{2+}$ substitution (ionic radius mismatch: $Cu^{2+}$ 0.82 Å vs. $Ni^{2+}$ 0.78 Å). Moreover, the LO peak exhibited a red shift from 577 to 516 $cm^{-1}$, indicating lattice strain or phonon softening resulting from Cu incorporation of Cu. The second-order modes (2TO and 2LO) also shifted to lower wavenumbers, reflecting changes in the phonon dispersion due to Cu-induced defects or substitutional incorporation into the NiO lattice. These shifts collectively confirm the successful doping of Cu and its influence on the local structure of the NiO matrix.

**Figure 4b** displays the photoluminescence (PL) spectra of undoped and Cu-doped NiO films, providing insight into electronic transitions and defect states. The undoped and Cu-doped NiO show strong emission bands at 564, 700, 749, 808, and 881 nm. The green band (564 nm) is associated with nickel vacancies ($V_{Ni}$) [29,30]. The origin of the red band (700 nm) may also be attributed to the oxygen vacancies ($V_O$). Upon Cu doping, the PL intensity significantly decreases, particularly in the 700–808 nm region. This quenching effect suggests that Cu atoms introduce non-radiative recombination centers, thereby reducing the radiative recombination process. In addition, peak broadening and red shifts in the emission bands indicate an increase in defect density and band tailing, likely due to Cu-induced structural disorder and new electronic states within the bandgap.

The strong photoluminescence peak at ~881 nm in NiO may be attributed to deep-level defect transitions within the bandgap. Because of its wide bandgap (~3.70 eV), emissions in the near-infrared region (~881 nm) are unlikely to stem from band-to-band recombination. Instead, such a peak is typically associated with radiative recombination involving defect states introduced by intrinsic imperfections. In NiO, these defects can include oxygen vacancies, nickel vacancies, complex defect clusters, or antisite defects. These defect centers form mid-gap es that facilitate electron-hole recombination far below the conduction and valence band edges, leading to the observed 881 nm emission. The pronounced intensity of this peak suggests a high density of such defects, which not only play a critical role in the optical properties of NiO but can also significantly influence its electrical behavior, an aspect important for gas sensor applications.

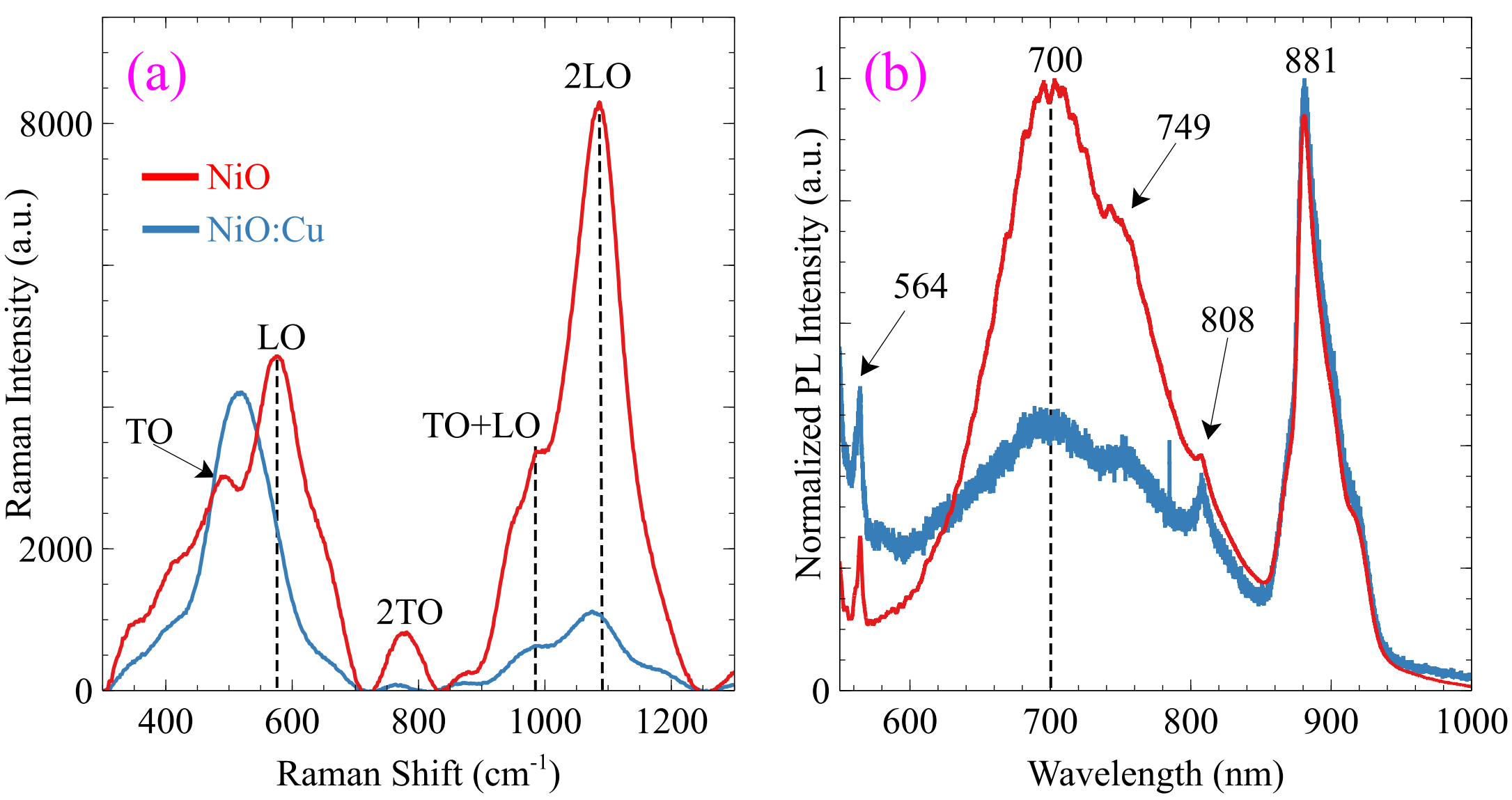

**Figure 4. (a) Raman and (b) photoluminescence analysis of NiO and NiO:Cu films on ITO-coated glass substrate.**

**Table 2. Peak analysis of Raman shifts for NiO and NiO:Cu films deposited on ITO-coated glass substrate.**

| Vibration Mode | Wavenumber ($cm^{-1}$) | |
|---|---|---|
| | NiO | NiO:Cu |
| TO | 493 | - |
| LO | 577 | 516 |
| 2TO | 776 | 766 |
| TO+LO | 988 | 983 |
| 2LO | 1089 | 1075 |

The comprehensive analysis of UV–vis, Raman, and photoluminescence spectra confirms that Cu doping significantly alters the structural and optical properties of NiO films. The UV–vis spectra reveal a slight reduction in the optical band gap from 3.70 eV (NiO) to 3.65 eV (NiO:Cu), indicating the introduction of intermediate energy levels within the band gap due to Cu incorporation. This is consistent with the PL spectra, which exhibit notable quenching and red shifts in the emission peaks, suggesting increased defect density and the presence of non-radiative recombination centers. The Raman analysis supports these findings by showing the disappearance of the TO phonon mode and red shifts in several vibrational modes, reflecting lattice distortion and enhanced defect formation. Together, these results highlight that Cu atoms effectively substitute for Ni atoms in the NiO lattice, inducing structural disorder and modifying the electronic transitions. This effect is advantageous for applications requiring enhanced surface reactivity and charge transport, such as gas sensing.

### 3.3 Electrical Properties

Figure 5 illustrates the current-voltage (I-V) and resistance-voltage (R-V) characteristics of Cu-doped NiO films under two different conditions: ambient air and exposure to the target gas. The films exhibit typical semiconducting behavior, with current increasing non-linearly as voltage increases. Under gas exposure $(I_G)$, the current significantly increases compared to ambient conditions $(I_A)$, suggesting that the presence of the gas reduces the resistance of the NiO film. At the bias voltage of 5 V and the presence of LPG, the current of NiO film increases from 249 mA $(I_A)$to 327 mA $(I_G)$, and the NiO:Cu film also increases from 265 mA $(I_A)$ to 480 mA $(I_G)$. The resistance of Cu-doped NiO films also decreases when the sample is exposed to an LPG environment. The $\Delta R$ values for NiO and NiO:Cu are 4 and 8 $\Omega$, respectively. This behavior is consistent with the p-type nature of NiO, where exposure to reducing gases (e.g., $H_2$, CO) donates electrons that neutralize holes, reducing the majority carrier density and altering the conductivity. The NiO:Cu film shows an even steeper increase in current under gas exposure, indicating enhanced sensitivity. Cu doping introduces additional active sites and defect levels that facilitate gas adsorption and charge transfer, thereby strengthening the interaction between the film surface and target gas molecules.

The sensitivity $(S)$ value of NiO and NiO:Cu thin films can be determined based on the following equation:

$$S = \frac{I_G}{I_A} \tag{6}$$

where $I_G$ (Ampere) is the current value when gas is flowing, and $I_A$ (Ampere) is the current at ambient condition. **Figure 6** presents the voltage-dependent sensitivity of pure NiO and NiO:Cu thin films. The sensitivity increases with applied voltage for both materials, but the trends and magnitudes are notably different. The red curve (NiO) shows a relatively low and slowly increasing sensitivity with voltage. This suggests that the charge carrier modulation or gas interaction at the surface is limited, and the NiO film exhibits a modest response throughout the measured voltage range. On the other hand, the blue curve (NiO:Cu) demonstrates a significantly higher sensitivity, especially in the intermediate to high voltage region. The sensitivity increases sharply with voltage and reaches a much higher saturation level compared to undoped NiO. This indicates that Cu doping dramatically enhances the film's ability to respond to gas molecules when an external bias is applied. It can be seen that at a bias voltage above 1 V, the NiO:Cu film has a higher sensitivity than that of NiO. At bias of 5V, the sensitivity is 1.3 and 1.8 for NiO and NiO:Cu. The voltage-sensitivity curve indicates that NiO:Cu outperforms pure NiO across all applied voltages, making it a more suitable candidate for bias-controlled gas sensing applications. The sharp rise and higher peak sensitivity also suggest that the optimal operating voltage range for NiO:Cu-based sensors can be tuned for improved performance.

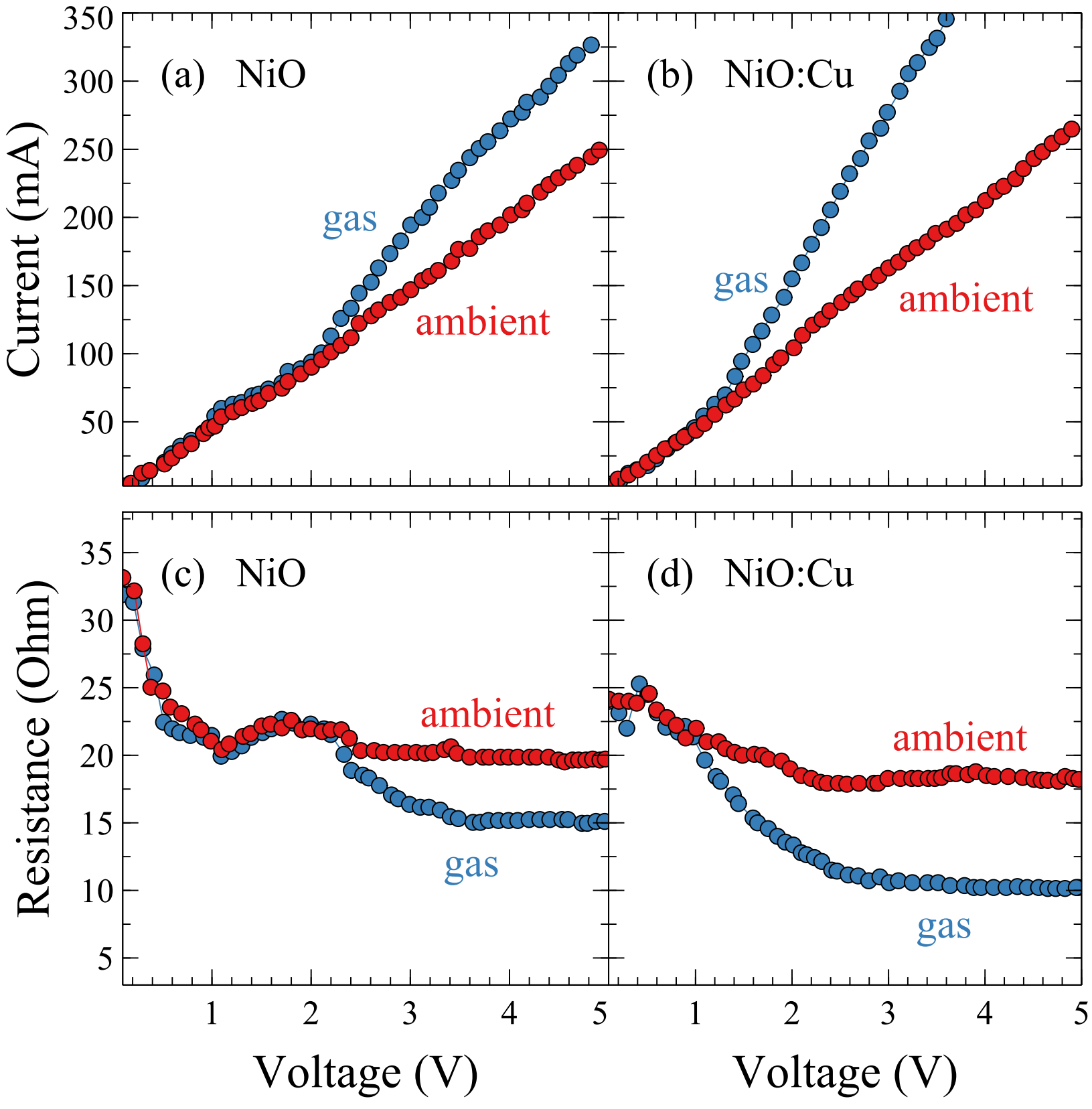

**Figure 5. I-V curves of (a) NiO and (b) Cu-doped NiO films grown by the spray pyrolysis technique on the ITO-coated glass substrate. Resistance as a function of applied voltage (0-5 V) for (a) NiO and (b) Cu-doped NiO films.**

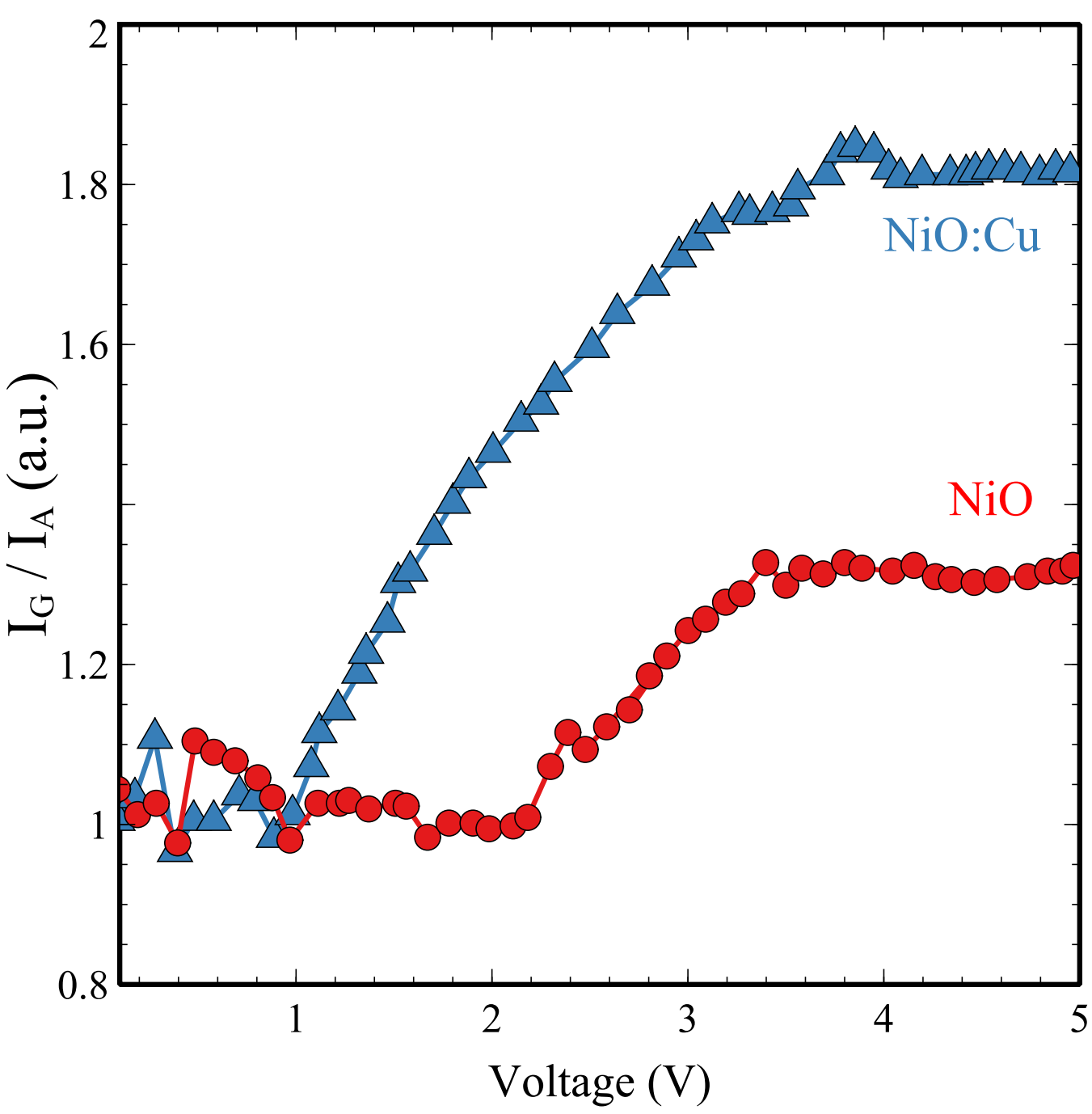

**Figure 6. Sensitivity as a function of voltage bias for NiO and NiO:Cu thin films grown by the spray pyrolysis technique.**

Overall, the data confirms that Cu doping significantly improves the gas sensing performance of NiO films by increasing carrier mobility, enhancing surface reactivity, and introducing additional pathways for charge exchange. These effects are attributed to the structural distortion and increased defect density evidenced by Raman and PL analyses. The synergistic effect of structural modification and electronic tuning makes NiO:Cu a more efficient sensing material compared to pristine NiO, particularly for applications requiring high sensitivity and low energy consumption.

## 4 Conclusion

In conclusion, Cu-doped NiO thin films synthesized via spray pyrolysis at 450 °C for 1.5 minutes exhibit significant improvements in structural, optical, and gas sensing properties compared to undoped NiO films. XRD and Raman analyses confirmed the successful incorporation of Cu into the NiO lattice, as evidenced by a reduction in crystallite size, lattice expansion, and vibrational mode shifts, indicating lattice distortion. The slight narrowing of the optical bandgap and the decrease in PL intensity upon doping suggest the formation of additional defect states, which facilitate charge carrier separation and reduce radiative recombination. These structural and optical modifications directly contribute to the enhanced gas sensing performance, where the Cu-doped films demonstrated a stronger and voltage-dependent sensitivity to LPG exposure. The observed improvement in sensing behavior is attributed to better charge transport, increased adsorption sites, and optimized surface reactivity resulting from Cu incorporation. Overall, this work highlights the potential of Cu-doped NiO thin films as efficient, cost-effective, and responsive materials for next-generation gas sensor devices.

## Funding Declaration

EN thanks the GBU45 research program from Institut Teknologi Sumatera for supporting this research (contract no. B/508/IT9.C/PT.01.03/2021).

## Data availability

The data that support the findings of this study are available on request from the corresponding author.

## Conflict of interest

The authors declare that they have no known competing financial interests or personal relationships that could have appeared to influence the work reported in this paper.

## CRediT Author Statement

**Eka Nurfani**: Conceptualization, Resources, Supervision, Funding acquisition, Writing - Original Draft. **Grace**: Investigation, Writing - Original Draft. **Mahardika Y. Darmawan**: Supervision, Funding acquisition, Writing - Review & Editing. **Resti Marlina**: Resources. **Jumaeda Jatmika**: Resources. **Asnan Rinovian**: Resources. **Aditya Rianjanu**: Resources, Funding acquisition.